# Current-induced domain wall motion in compensated ferrimagnet

Saima A Siddiqui[1], Jiahao Han[1], Joseph T Finley[1], Caroline A Ross[2] and Luqiao Liu[1]

[1]Department of Electrical Engineering and Computer Science, Massachusetts Institute of Technology, Cambridge, MA 02139

[2]Department of Materials Science and Engineering, Massachusetts Institute of Technology, Cambridge, MA 02139

Due to the difficulty in detecting and manipulating magnetic states of antiferromagnetic materials, studying their switching dynamics using electrical methods remains a challenging task. In this work, by employing heavy metal/rare earth-transition metal alloy bilayers, we experimentally studied current-induced domain wall dynamics in an antiferromagnetically coupled system. We show that the current-induced domain wall mobility reaches a maximum close to the angular momentum compensation. With experiment and modelling, we further reveal the internal structures of domain walls and the underlying mechanisms for their fast motion. We show that the chirality of the ferrimagnetic domain walls remains the same across the compensation points, suggesting that spin orientations of specific sub-lattices rather than net magnetization determine Dzyaloshinskii-Moriya interaction in heavy metal/ferrimagnet bilayers. The high current-induced domain wall mobility and the robust domain wall chirality in compensated ferrimagnetic material opens new opportunities for high-speed spintronic devices.

Antiferromagnetic materials have fast intrinsic magnetization dynamics and are insensitive to magnetic fields, making them potential candidates for the next generation of dense, high-speed spintronic devices [1-5]. However, their magnetic state is difficult to manipulate and detect electrically. Therefore, studying their high frequency switching dynamics using electrical methods remains challenging. In contrast, ferrimagnetic materials, many of which have antiparallel aligned sublattices, provide another possible platform for realizing fast device operation [6,7], with the advantage that their magnetization state can be detected or altered even near the compensation point because their sensitivity to current-induced spin torque and their magneto-electric [8-12] or magneto-optical [13,14] response does not disappear. Rare earth-transition metal (RE-TM) alloys are well known ferrimagnets in which the RE and the TM sublattices align antiparallel with each other reducing the net angular and magnetic moments. High frequency magnetic resonance and picosecond magnetic switching by optical pulses have been demonstrated in thin films of these materials [6,7,15,16], motivating their application in ultrafast spintronic devices. Recently it was demonstrated that electrical current could be used as an efficient switching mechanism for RE-TM ferrimagnets even at the compensation point via spin-orbit torque [8-12]. However, limited by the quasi-static measurement techniques, the electrically driven switching dynamics in these materials is yet to be explored. Experiments on current-induced domain wall (DW) motion not only provide a convenient way to study the time-dependent switching dynamics of multi-domain magnets [17-19], but also can lead to useful high density memory and logic devices [20,21]. Very recently, it was shown using magnetic field driven experiments that angular momentum compensated ferrimagnetic materials possess great velocity advantages [22], which provides the possibility of reaching high operational speed in devices based on those materials. However, it remains unclear if the same speed maximum is retained in current-induced DW motion. Particularly, different from field driven case, additional factors such as Dzyaloshinskii-Moriya interaction (DMI) and the domain wall chirality play important roles in current-induced experiment [23-25]. It is an open question how DMI evolves in the presence of two oppositely aligned sublattices, and whether it supports the needed chirality for efficient spin orbit torque induced DW motion at the compensation points. To answer these questions, we experimentally study the fast current-induced DW dynamics in compensated ferrimagnets by characterizing the DW motion in Pt/Co$_{1-x}$Tb$_x$ wires with various chemical compositions and reveal the physical mechanisms behind the phenomena.

A series of Pt (3 nm)/Co$_{1-x}$Tb$_x$ (2-3 nm)/SiN$_x$ (3 nm) samples was deposited using magnetron sputtering. The Pt underlayer provides the spin-orbit torque while SiN$_x$ is used as an insulating capping layer. Figure 1(a) shows the coercive fields ($H_c$) and the saturation magnetizations ($M_s$) of unpatterned films at different compositions. Because of the large bulk perpendicular magnetic anisotropy, the easy axes of the samples are oriented out-of-plane. Co$_{1-x}$Tb$_x$ reaches its magnetization compensation point at $x = 0.34$, where $M_s$ is minimum and $H_c$ diverges. We note that this compensation composition for Pt/Co$_{1-x}$Tb$_x$ is slightly different from what has been observed previously for Ta/Co$_{1-x}$Tb$_x$ samples [8], probably due to the extra contribution from the proximity-induced magnetization in Pt [26]. Due to the unequal gyromagnetic ratios of RE and TM elements, the concentration where the magnetization reaches compensation is different from that with zero total angular momentum. Using the literature values of $g$ factors of Co (~2.2) [27] and Tb (~1.5) [28,29] atoms, we can also estimate that the angular momentum compensation point is around $x \approx 0.25$.

The deposited films were patterned into micron size wires and Hall bar structures for DW motion measurement and anomalous Hall resistance ($R_{AH}$) characterization, respectively. The schematic structure of the Hall bar is shown in Fig. 1(b) along with the measurement set-up. $R_{AH}$ changes sign between Co$_{0.67}$Tb$_{0.33}$ and Co$_{0.64}$Tb$_{0.36}$ [Fig. 1(c)], which is consistent with a transition from being Co-dominated to being Tb-dominated in magnetic moment [8,30]. Figure 2(a) shows the schematic of the set-up for studying DW motion in 2-5 $\mu$m wide Co$_{1-x}$Tb$_x$ magnetic wires. All the velocity measurements are done at room temperature. First, a large external magnetic field along the out-of-plane $z$ direction ($H_z$) was applied to saturate the magnetization. Next, a 1~100 ms duration magnetic field pulse in the opposite direction was applied to nucleate and initiate DW propagation from the large contact pad region. A magneto-optical Kerr effect (MOKE) microscope was utilized to track the position of the DWs. By sweeping $H_z$ on a series of samples, we verified that the yellow and green regions in our MOKE image represent domains where the Co sublattice points along the -$z$ and +$z$ direction, respectively, for all chemical compositions studied. This is consistent with the observations that the TM sublattice dominates the Kerr signal for TM-RE alloys in the visible light regime [14,15]. For convenience, we will label the two different domains as ↓ and ↑ domains in the following discussion, where ↓ and ↑ denote the orientations of the Co sublattice. After the initial nucleation of DWs, electrical current pulses of 20-ns duration were then applied to the wires to move the DW. The initial and the final positions of the DWs were measured with MOKE microscopy and the velocity was calculated from the DW displacement and the total pulse length. Figure 2(b) gives an example of the positions of DWs after multiple current pulses of the same width. Samples with different channel widths (2 $\mu$m - 5 $\mu$m) were tested, but no dependence of velocity on sample width was observed (see Supplementary Fig. S1 [31]).

The DW velocity as a function of the applied current density in a series of Pt/Co$_{1-x}$Tb$_x$ wires ($x$ = 0.17 - 0.41) is summarized in Fig. 2(c). In our experiment, both ↑↓ and ↓↑ DWs move along the direction of the charge current (see Supplementary Fig S2 [31]), similar to the direction of DW motion observed previously in Pt/ferromagnetic systems [23,24], which supports the essential role of spin-orbit torque. To exclude the contributions from differences in threshold current densities between samples, we focus on the regions where the velocity and the current density roughly satisfy a linear relationship. The DW mobility of all the samples, defined as the ratio between the velocity and the current density, is determined by fitting the slopes of the linear regions in Fig. 2(c). The results are summarized in Fig. 2(d) and show that the DW mobility varies by more than an order of magnitude depending on the chemical composition. Starting from the Tb-dominant samples, the mobility increases with Co concentration and reaches a maximum around $x \approx 0.21 \sim 0.26$, which agrees with the estimated angular momentum compensation point range, considering the error bars in our experimental data and the angular momentum compensation calculation. We note that the fact that DW attains its maximum speed when the net angular momentum rather than the magnetization reaches zero is also consistent with recent study on the field-induced DW motion [22], despite different driving forces.

The DW mobility [~ $5 \times 10^{-10}$ m$^3$/(A·s)] obtained in our compensated CoTb sample is much higher than what was observed previously with ferromagnetic layers (e.g., CoFeB [23,32] and Co/Ni/Co [24]) where the mobility is $0.2 \times 10^{-10}$ m$^3$/(A·s) – $1 \times 10^{-10}$ m$^3$/(A·s), and is comparable to the values found in synthetic antiferromagnet multilayers [33]. Compared with these previous experiments, the current densities used in our experiment are relatively low (< $5 \times 10^{11}$ A/ m$^2$ *vs* 1~$5 \times 10^{12}$ A/ m$^2$). To achieve even higher absolute values of DW velocity, larger current densities are required. We found that above a certain current density (defined as the maximum current density), nucleation of new domains starts to occur, which puts an upper limit on the applicable current. This is similar to a previous observation in the Ta/CoFeB/MgO system, where the breakdown of DW motion was attributed to current-induced weakening of the perpendicular anisotropy [32]. The decrease in anisotropy was observed in the temperature-dependent vibrating sample magnetometry (see Supplementary Fig. S3 [31]). It is also noted that while magnetic anisotropy decreases rapidly due to heating effect, the changes of magnetization remains small (less than 10% of room temperature value) before the films lose coercivity (Supplementary Fig. S3(b) [31]), suggesting that the drift of magnetization is insignificant during the current pulse application.

To understand the evolution of DW velocity as a function of net moment, we modelled the DW motion for ferrimagnetic materials with two unequal sublattices. Previously it was demonstrated that the magnetic dynamics of ferrimagnets could be described with the Landau–Lifshitz–Gilbert equation by replacing the regular gyromagnetic ratio and damping coefficient with the effective values, $\gamma_{eff}$ and $\alpha_{eff}$ : $\frac{d\hat{m}}{dt} = -\gamma_{eff}\hat{m} \times \vec{H} + \alpha_{eff}\hat{m} \times \frac{d\hat{m}}{dt} - \gamma_{eff}\frac{\hbar\theta_H j}{2e\mu_0 M_{eff} t}(\hat{m} \times \hat{\sigma} \times \hat{m})$ . Here, $M_{eff} = M_1 - M_2$, , $\gamma_{eff} = (M_1 - M_2)/(S_1 - S_2)$, and $\alpha_{eff} = (\alpha_1 S_1 + \alpha_2 S_2)/(S_1 - S_2)$ with $M_{1,2}$, $S_{1,2}$ and $\alpha_{1,2}$ representing the magnetization, angular moment per unit volume and damping coefficient of the two sublattices [6,7].

$\hat{m}$ denotes the unit vector along the direction of $\vec{m}_1 - \vec{m}_2$ (Néel vector). $t$, $\hat{\sigma}$, $\theta_H$ and $j$ are the thickness of the magnetic film, the orientation of spins injected into the ferrimagnet, spin Hall angle and applied current density, respectively. As is shown in Supplementary Note 3 [31], under this replacement of $\gamma_{eff}$ and $\alpha_{eff}$, the DW velocity of a ferrimagnetic wire can be derived as: $v = v_S / \sqrt{1 + (j_S/j)^2}$, where $v_S = \gamma_{eff} \Delta H_D$ and $j_S = \frac{4e\mu_0 \alpha_{eff} M_{eff} t}{\pi \hbar \theta_H} H_{DMI}$ represent the saturation velocity and saturation current density, respectively. Here $H_{DMI}$ is an in-plane effective field, originating from DMI (see discussions below) and $\Delta$ is the domain wall width. This expression of DW velocity is similar to that of the ferromagnetic systems except that $\gamma_{eff}$ and $\alpha_{eff}$ are utilized [25]. It can be seen that unlike a typical ferromagnet whose highest DW velocity is limited by the chirality stabilizing force -- the DMI effective field, a ferrimagnet does not have a speed limit because $\gamma_{eff}$ and $\alpha_{eff}$ diverge at the compensation point. This ensures the linear relationship between $j$ and $v$ exists throughout the whole current range. DW velocities in ferrimagnets with different net angular moment are calculated and compared in Fig. 2(e). It shows that the compensated ferrimagnetic material has velocity advantages at large or intermediate current densities, which is consistent with our experimental observations.

DMI plays an important role in stabilizing the DW chirality and alleviating the velocity reduction caused by Walker breakdown. In a ferromagnet, the effective field from DMI directly determines the highest DW velocity that can be reached [25]. For a compensated ferrimagnet, as discussed above, the DW velocity is no longer restrained by $H_{DMI}$. However, a non-zero DMI is still critical to ensure that a Néel type of DW is favorable, for which the spin orbit torque has the highest efficiency (Supplementary Note 3 [31]). So far little is known about the DMI at the interface between heavy metals and ferrimagnetic alloys. In particular, it is not clear how the DW chirality varies when the net magnetization or net angular momentum goes through zero as the composition varies. To characterize DMI in ferrimagnetic $Co_{1-x}Tb_x$, we measured current-induced DW velocities as a function of in-plane field ($H_x$) along the wire direction. The results from a $Co_{0.79}Tb_{0.21}$ sample are illustrated in Fig. 3(a) and 3(b). It can be seen that under positive (negative) $H_x$, the ↓↑ DW in $Co_{0.79}Tb_{0.21}$ moves faster (slower) compared with the zero field case. The trend is opposite for ↑↓ DWs, and the DW velocities even change direction at $H_x = \pm 1000$ Oe. The fact that the motion for one type of DW is enhanced while the other is suppressed is consistent with the Néel wall characteristics, where the applied $H_x$ strengthens (or weakens) the effective DMI field [Fig 3(c)]. This is in contrast with other DW configurations (e.g., a Bloch wall), where a symmetric variation of the DW velocity under $H_x$ is expected.

To answer the question of whether the DW changes its chirality at the compensation points, we plot the dependence of DW velocity as a function of $H_x$ in Fig. 3(e)-3(g) for three different $Co_{1-x}Tb_x$ samples. Since the angular momentum and magnetization compensation points are at $x = 0.25$ and $0.34$ respectively, the samples with $x = 0.21$, $0.33$, and $0.41$ in Fig. 3(e)-3(g) represent three different cases: Co-dominant in both angular momentum and magnetization, Co-dominant in magnetization and Tb-dominant in angular momentum, and Tb-dominant in both angular momentum and magnetization, respectively. First, we find that there is no qualitative change in the DW motion characteristics across the angular momentum compensation point [Fig. 3(e) and 3(f)]. Under an $H_x$ field, the $Co_{0.67}Tb_{0.33}$ sample exhibits similar behavior to the previously discussed $Co_{0.79}Tb_{0.21}$ sample, where the velocity of ↓↑ (↑↓) DWs increases (decreases) under small positive $H_x$. However, across the magnetization compensation point, the opposite trend was seen [Fig. 3(e)], where the velocity of the ↓↑ (↑↓) DWs decreases (increases) under the same positive $H_x$ field. The sign reversal in the $v$ vs $H_x$ slopes across the magnetization compensation point can be explained by the schematic DW structures shown in Fig 3(c) and 3(d). A positive $H_x$ field will stabilize the ↓↑ DW in magnetically Co-dominant samples, while it will destabilize the ↓↑ DW in Tb-dominant samples. Therefore, the sign reversal reflects that the left-handedness is maintained throughout all our $Pt/Co_{1-x}Tb_x$ samples, suggesting that *the DMI is correlated with the spin orientations of specific sub-lattices rather than the net magnetization*. The in-plane magnetic fields which overcome DMI and result in zero domain wall velocity for the above three compositions are summarized in Fig. S5 [31]. It is noted that $H_{DMI}$ does not simply scale following the expected relationship of $H_{DMI} =$ $D/M_{eff}t\mu_0\Delta$, where $D$, $t$, $\mu_0$ and $\Delta$ representing the interfacial DMI energy density, film thickness, vacuum permeability and DW width [25]. Instead, samples with higher Tb concentration tend to have larger $H_{DMI}$, which could be attributed to the strong spin orbit coupling associated with rare earth element. We note that besides allowing for fast DW movement, the strong DMI and robust chirality exhibited in our compensated ferrimagnet provide the possibility of engineering skyrmion structures with zero total angular momentum. Because of the cancelation of the side deflections from the skyrmion Hall effect of two sublattices, these compensated skyrmions are expected to have greatly enhanced mobility [4,34].

To summarize, we experimentally investigated the fast domain wall dynamics in $Co_{1-x}Tb_x$ ferrimagnetic samples. We found that the domain wall mobility reaches a maximum in samples close to compensated angular momentum and it is higher than those in the ferromagnetic electrodes. The high domain wall velocity in a compensated ferrimagnetic material is consistent with our theoretical modelling, where it is shown that the absence of velocity saturation ensures a high mobility. By measuring the influence of in-plane field on the domain wall velocity, we further demonstrated that the domain walls have chiral internal structures which are stabilized by the Dzyaloshinskii-Moriya interaction and the same chirality is maintained across the compensation points. Thus we identifies that the Dzyaloshinskii-Moriya interaction in ferrimagnetic materials is related to the spins of the sublattices in contrast to the net magnetization. Our study on current-induced domain wall motion in ferrimagnets opens the opportunity to electrically probe the fast domain wall dynamics in angular-momentum compensated systems. The low magnetic moment, large electrical and optical response, as well as the possibility of reaching high speed dynamics makes it highly attractive to employ ferrimagnets for spintronic applications.

This research was partially supported by the National Science Foundation under grant 1639921, and the Nanoelectronics Research Corporation (NERC), a wholly-owned subsidiary of the Semiconductor Research Corporation (SRC), through Memory, Logic, and Logic in Memory Using Three Terminal Magnetic Tunnel Junctions, an SRC-NRI Nanoelectronics Research Initiative Center under Research Task ID 2700.001.

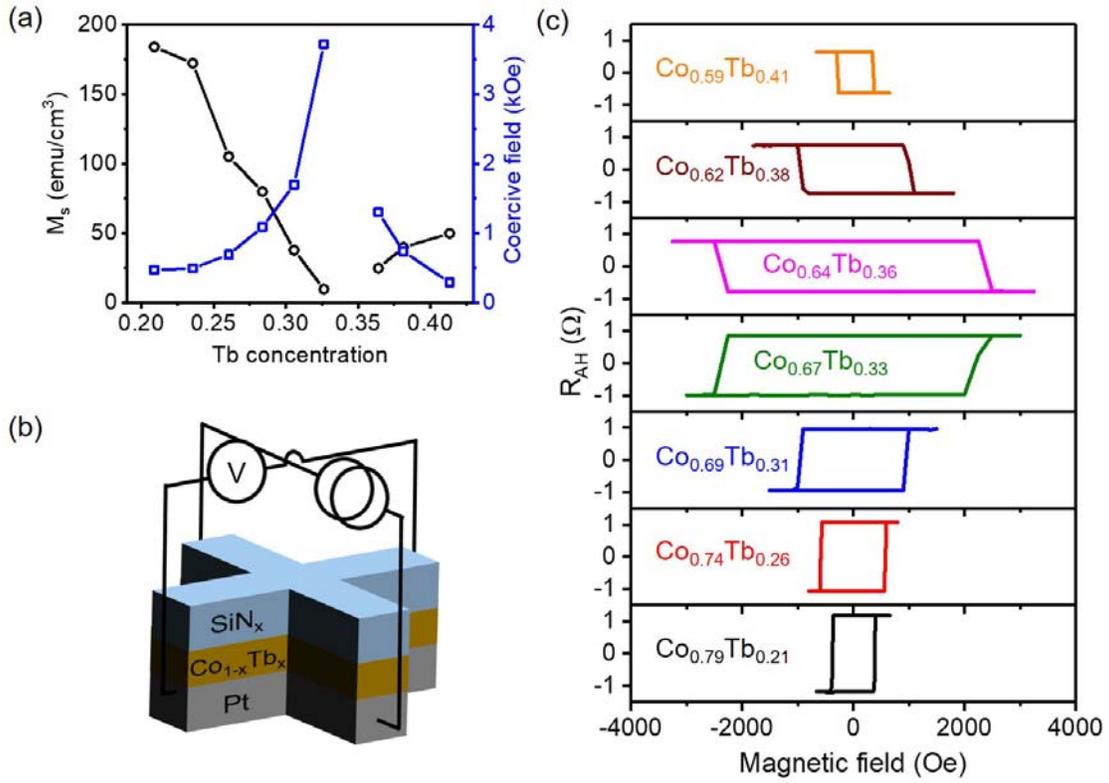

FIG. 1. (a) Saturation magnetization and coercive fields of $Pt/Co_{1-x}Tb_x$ thin films from vibrating sample magnetometry. (b) Schematics of the device geometry and electrical setup for Hall resistance measurement. (c) Anomalous Hall resistance of 4-$\mu$m wide $Pt/Co_{1-x}Tb_x$ Hall bars. The coercive fields of the patterned structures differ from that of the continuous films due to domain nucleation and pinning processes at wire edges.

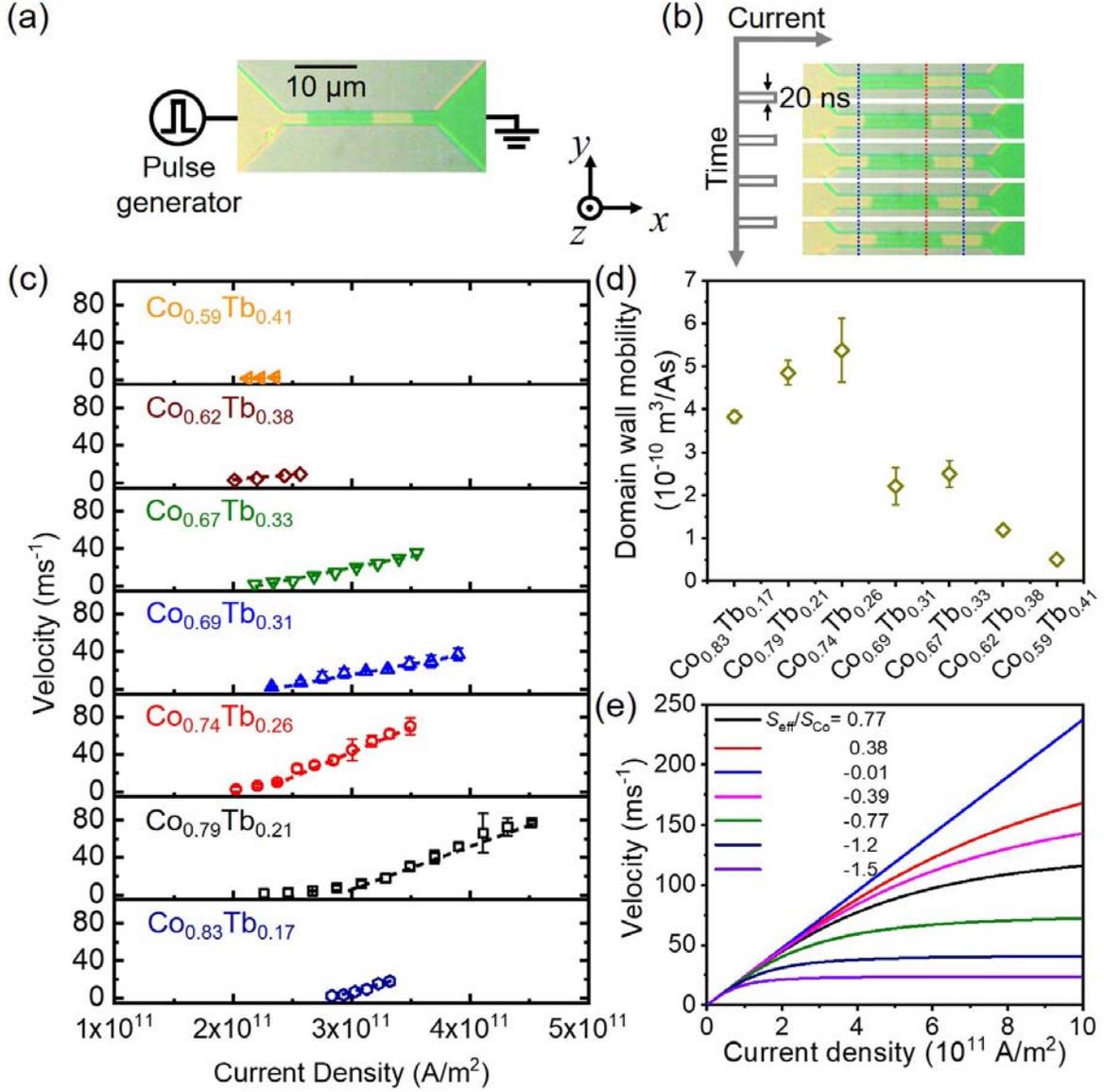

FIG. 2. (a) Electrical set-up for the domain wall motion measurement. The yellow and the green regions represent ↓ and ↑ domains in the MOKE microscope image of 4-$\mu$m wide $Co_{0.69}Tb_{0.31}$ wire. (b) Domain wall motion in $Co_{0.69}Tb_{0.31}$ wire with consecutive current pulses. Blue and red dotted lines show the initial positions of ↓↑ and ↑↓ domain walls in the top MOKE image, respectively. (c), Domain wall velocity as a function of current density for Pt/$Co_{1-x}Tb_x$ at $x$ = 0.17, 0.21, 0.26, 0.31, 0.33, 0.38 and 0.41 (from bottom to top panel). The error bars reflect standard deviations from multiple measurements. (d) Domain wall mobility extracted from the dotted lines in (c) for Pt/$Co_{1-x}Tb_x$. (e) Calculated current-induced domain wall velocity for a series of ferrimagnetic samples with different net angular momentum, $S_{eff}$.

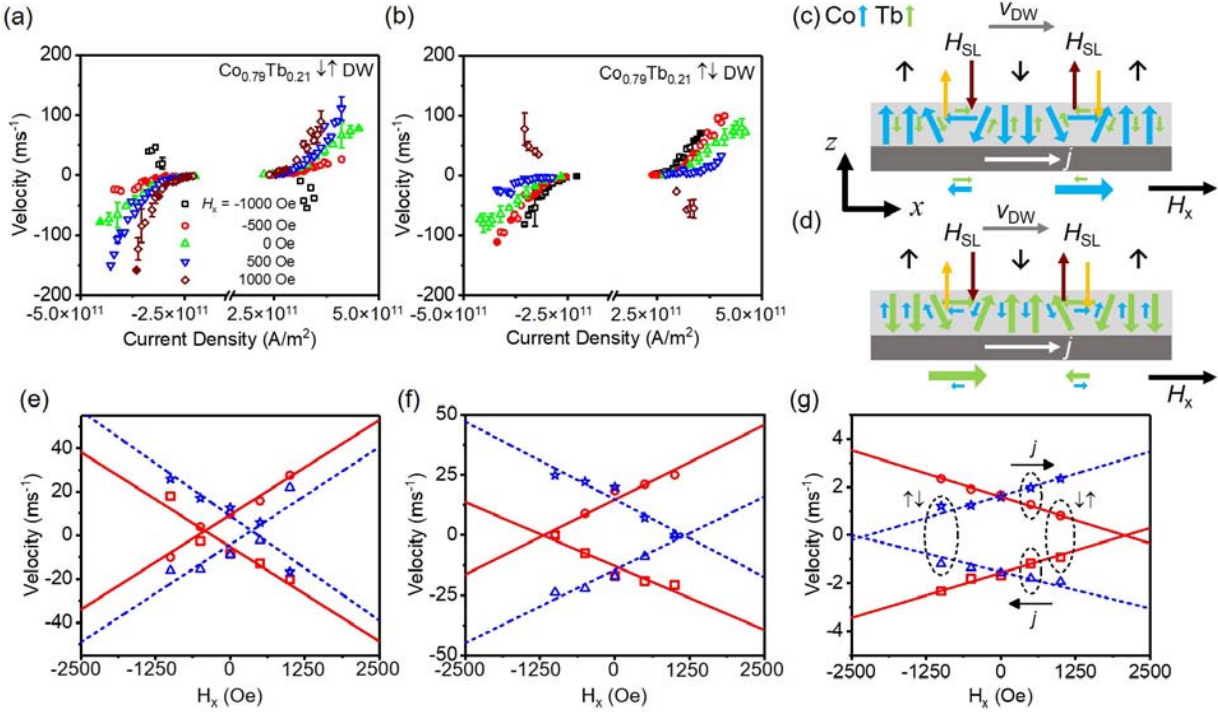

FIG. 3. Domain wall velocity as a function of current density at different longitudinal fields along the length of the Pt/Co$_{0.79}$Tb$_{0.21}$ sample for (a) ↓↑ and (b) ↑↓ domain walls. Schematic illustration of the domain wall texture for ↓↑ and ↑↓ Néel domain walls in (c) magnetically Co-dominant and (d) magnetically Tb-dominant samples. Blue and green arrows represent the magnetizations from Co and Tb sublattices, respectively. The chirality of the ↓↑ and ↑↓ domain walls remains the same as the composition changes from the Co- dominant to Tb-dominant side. The effective fields from Slonczewski-like torque ($H_{SL}$) on Co and Tb sublattices are shown by yellow and brown arrows, respectively. It can be seen that $H_{SL}$ on the two sublattices work constructively to move domain walls. The domain wall motion direction remains the same in both samples. The black ↓ and ↑ arrows show the domain orientation as detected in the MOKE measurement. The length of the blue and green arrows below the domain wall region reflects the influence of external field $H_x$ on domain wall chirality. Domain wall velocity as a function of in-plane field for samples of (e) Pt/Co$_{0.79}$Tb$_{0.21}$, (f) Pt/Co$_{0.67}$Tb$_{0.33}$ and (g) Pt/Co$_{0.59}$Tb$_{0.41}$ wires, respectively. Red squares (negative current) & red circles (positive current) represent ↓↑ domain walls and blue triangles (negative current) & blue stars (positive current) represent ↑↓ domain walls, respectively. Red solid lines and blue dashed lines are the linear fit of the experimental data for ↓↑ and ↑↓ domain walls. There is a sign reversal in the slopes of the red and blue lines between (e), (f) and (g).